# Reversible Data Hiding in Encrypted Images using Local Difference of Neighboring Pixels

Ammar Mohammadi, and Mansor Nakhkash

*Abstract*— This paper presents a reversible data hiding in encrypted image (RDHEI), which divides image into non-overlapping blocks. In each block, central pixel of the block is considered as leader pixel and others as follower ones. The prediction errors between the intensity of follower pixels and leader ones are calculated and analyzed to determine a feature for block embedding capacity. This feature indicates the amount of data that can be embedded in a block. Using this pre-processing for all blocks, we vacate rooms before the encryption of the original image to achieve high embedding capacity. Also, using these features, embedded data is extracted and the original image is losslessly reconstructed at the decoding phase. Comparing to existent RDHEI algorithms, not only embedding capacity is increased by the proposed algorithm, but also perfect reconstruction of the original image is realized by content owner without having data hider key. Experimental results confirm that the proposed algorithm outperforms state of the art ones.

## I. INTRODUCTION

Reversible data hiding (RDH) intends to embed a secret data in a cover signal like image, while cover signal is reconstructed absolutely after perfect extraction of the embedded secret data. The most RDH schemes in plaintext image use correlation of neighboring pixels to embed secret data [1-5]. The RDH has many applications, among them are copyright protection, authentication, cover communication, and so on. The RDH in encrypted image (RDHEI) is also used in cloud computing to preserve content owner privacy and provide a platform to embed additional data in encrypted image including some necessary notifications, content owner identification, source and destination information and so on. In RDHEI, without knowing original content of the image or encryption key, additional data may be embedded in encrypted image. At the decoding phase, original image and embedded data can be restored and extracted respectively.

In recent years many research papers have been devoted to RDHEI [6]. Generally, they can be classified into three groups: reserving room before encryption (RRBE), vacating room by encryption (VRBE) and vacating room after encryption (VRAE). Also, the RDHEI methods vacate room in encrypted image based on two fundamental approaches: compressing encrypted image [7, 8] and exploiting the correlation between neighboring pixels [8-23]. In another classification, RDHEI schemes may be categorized to separable or joint methods. In separable methods, the data extraction and image decryption are performed separately, whereas the data extraction cannot be realized without knowing the information of decrypted image in joint methods.

The idea of compressing the encrypted image is introduced in [24, 25]. It is obvious that compressing the encrypted image vacates room to embed secret data. Using this idea, [7, 8] perform RDHEI by VRAE that may be more practical than two other ones; however, high embedding capacity is not achievable in their scheme. Moreover, RDHEI scheme in [7] cannot reconstruct the original image losslessly under some circumstances. This scheme is improved by [8] employing the correlation between neighboring pixels and distributed source coding. At the decoding phase, message extraction is separately done in both of them.

In most RDH scheme, the correlation of neighboring pixels is exploited to calculate prediction errors of the original image and in turn, it will be used to embed the secret data. The correlation is, also, used in RDHEI to perform data hiding. In 2013, Reference [11] proposed a scheme of commutative reversible data hiding and encryption. Data hider first fragments the original image into a series of non-overlapping blocks. To encrypt original image, intensities of two neighboring pixels are masked by the same pseudo-random bits. In this way, the difference of the neighbors is not changed and thus, the spatial redundancy in encrypted image is preserved to embed secret data. There is a pseudo-random permutation process in this scheme to prevent some possible attacks that may occur because of information leakage and no encryption of the difference of neighboring pixels. In 2014, Zhang *et al*. [13] presented a RDHEI that instead of embedding data in encrypted images directly, some pixels are predicted before encryption so that the secret data can be embedded in the prediction errors. Thus, a part of these prediction errors will not be encrypted. Another attempt in [12] introduces a RDHEI, in which the cover image is divided into separate blocks and multigranularity encryption is applied to attain the encrypted image by random permutation of blocks and random permutation of pixels in each block. It preserves the correlation of neighboring pixels in each block that is used to calculate prediction errors. In 2016, Xu and Wang [15] reported a RDHEI using correlation of sample pixels and non-sample ones. Sample pixels as reference points are used to calculate prediction errors of non-sample pixels. A stream cipher is used to encrypt sample pixels and a specific encryption procedure is planned to encrypt prediction errors of non-sample ones. Accordingly, a part of prediction errors that are more frequent is not encrypted. This procedure may be organized using a modified version of histogram shifting and difference expansion technique. In this scheme, a part of prediction errors will not be encrypted. Huang *et al.* [14] also propose a new framework for RDHEI. Their new framework allows the numerous RDH schemes directly be exploited in the encrypted domain. They divide image into different blocks and use the correlation of neighboring pixels in each block to embed secret data. They encrypt whole pixels in a block by using a similar random integer. Thus, it will preserve correlation of pixels in a block that makes redundancy to embed secret data in encrypted image. Although perservering the correlation of neighboring pixels makes possibility to embed data in an encrypted block, it may lead to an effective know/chosen-plaintext attack. Using block permutation, they try to protect against this attack by random distribution of blocks in the whole image. In 2019, Yi an Zhou [19] proposed a high embedding capacity scheme that exploits correlation of pixels in a block to embed secret data in the encrypted image. Their scheme includes four main steps, image blocking, pixel grouping, pixel labeling, and payload embedding. In this scheme, the image encryption involves two procedures: block permutation and pixel modulation. All pixels in a selected block are added by a similar random integer to organize pixel modulation. In this procedure, the correlation of neighboring pixels in blocks is preserved to embed secret data. Its drawback is that the difference of pixels in a block will not be private because a similar random integer is added to whole pixels in a block. This leads to information leakage in the encrypted image. However, security analysis has confirmed the robustness of this scheme in withstanding brute-force and know/chosen-plaintext attacks. This scheme significantly improves hiding capacity in comparison with previous ones.

All five aforementioned methods use VRBE approaches. For the decoding phase, these schemes are separable. Original image perfectly is reconstructed in these schemes. All of them do not encrypt difference of some pixels in the original image because of exploiting correlation of the neighboring pixels to embed data in encrypted image. Some attacks may be possible in these schemes because of information leakage.

In 2011, Zhang [9] proposed a RDHEI in accordance with VRAE procedure so that secret data is embedded in an image by modifying a small proportion of encrypted image. They take advantages of spatial correlation in natural image to extract data. This scheme is a joint one. Data extraction and image reconstruction perfectly will be done just under some circumstances. This scheme was later improved by Hong *et al.* [10] using spatial correlation between neighboring blocks and a side-match technique. In 2016, Zhou *et al.* [17] improved two previous schemes in terms of hiding capacity and accuracy of data extraction and original image recovery. They exploit several binary public keys to embed data that are selected according to a criterion of maximizing the minimum Hamming distance among all keys. At the decoder side, a powerful two-class SVM classifier is exploited to separate encrypted and non-encrypted image patches. The reported methods in [9], [10] and [17] use VRAE approaches. At the decoding phase, these schemes are joint and the original image is perfectly reconstructed just under some circumstances.

Using a traditional RDH method, Ma *et al.* [20] realize a RDHEI that reserves room before encryption by embedding LSBs of some pixels into other ones. After encryption, the positions of these vacated LSBs are used to embed data. The scheme proposed in [21] improves that of [20] using patch-level sparse representation. Also, Puteaux and Puech [18] present two high capacity RDHEI schemes by RRBE. In one of them, perfect reconstruction of the original image is guaranteed. Their scheme employs the correlation of two neighboring pixels so that a pixel can be predicted by adjacent one. Therefore, prediction errors may be calculated in plaintext image that are used to make a location binary map. After encryption, the data can be embedded in MSBs of encrypted pixels using this location binary map. Yin *et al.* [23] employ based idea of [18] to reserve room before encryption and improve this paper using source coding. They use median edge detector (MED) predictor to label each pixel to reserve more significant bits of image pixels, instead of just MSB. These labels will be enormous in size so they apply Huffman coding to losslessly compress them.

Discussed RRBE schemes guarantee the perfect original image reconstruction at decoding phase and they are separable.

This paper reports a RDHEI scheme that is the extension of the idea reported in [18]. It uses local difference of separated blocks of the original image to compute prediction errors and obtains a feature for each block analyzing the prediction errors. Such a feature shows the amount of bits embedded in the blocks and we call it block capacity feature (BCF). The BCFs are employed to embed both themselves and secret data in an encrypted image. At the decoding phase, at first, the BCFs are hierarchically extracted and then, using these BCFs, secret data can be extracted and original image may be reconstructed separately. Our method improves the embedding capacity significantly in comparison with that of [18]. Implementing of our proposals and comparing the results with those of the existent RDHEI algorithms, we demonstrate the improvement in embedding capacity.

## II. PROPOSED SCHEME

In this section, we present our proposed scheme in details that comprise analyzing hiding capacity, reserving room before encryption, encrypting the original image, embedding BCFs and secret data, extracting data and recovering the original image.

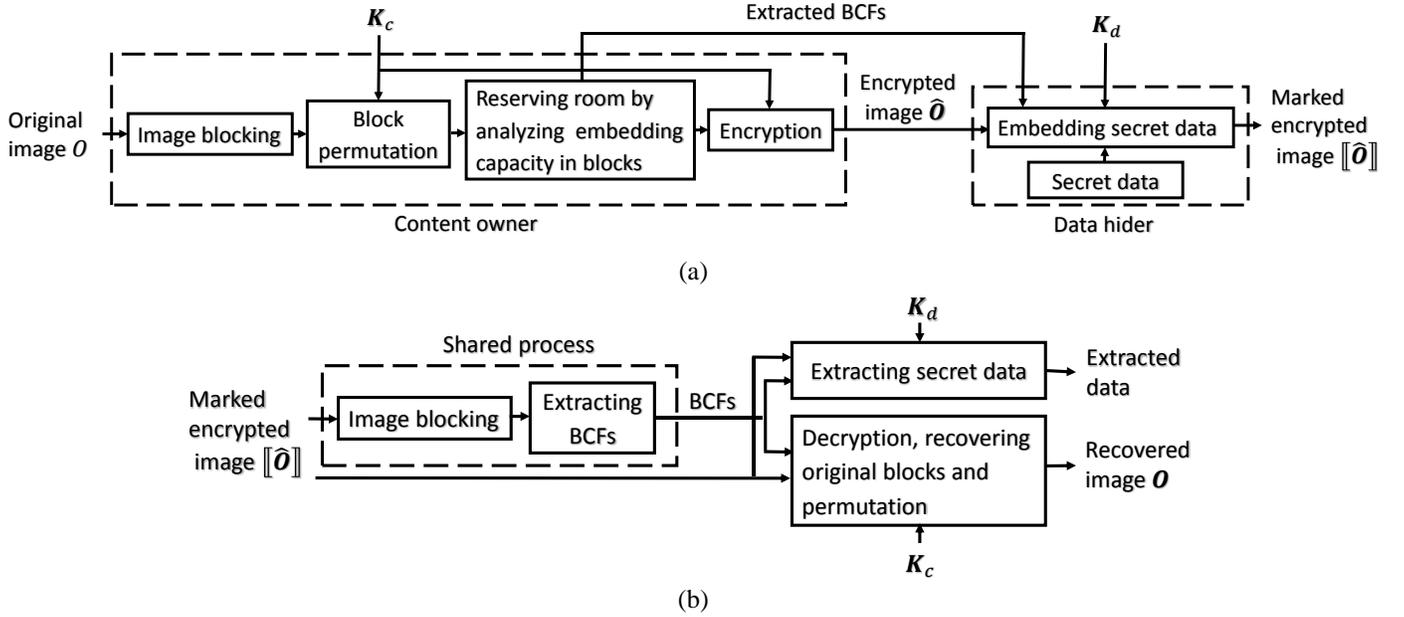

Fig. 1. Block diagram of the proposed scheme. (a) Embedding data and forming marked encrypted image. (b) Extracting data and recovering original image.

Fig. 1 indicates an overview of the proposed scheme. Fig. 1a demonstrates the procedure for embedding secret data in encrypted image that leads to construct the marked image and Fig. 1b illustrates the data extraction process and the recovery of the original image. According to Fig. 1a, in order to embed data, the image is first divided into non-overlapping blocks. These blocks are permuted with content owner key ($K_c$). Afterwards, pre-processing the permuted blocks, the BCF of each block is determined and the permuted image is encrypted. The encrypted image is obtained using bitwise XOR of the permuted one with a stream cipher that is created by an encryption algorithm with an input secret key ($K_c$). Both permutation and bitwise XOR should be considered as the encryption of the original image. Data hider embeds both the BCFs and the secret data in the encrypted image employing BCFs in a hierarchical procedure. He/She can use his/her own key ($K_d$) to encrypt the secret data. The resultant is the marked encrypted image.

At the decoder side, as shown in Fig. 1b, the data extraction and lossless reconstruction of the original image individually can be performed. According to this figure, the image is first divided into non-overlapped blocks with the same size in encoding procedure. Then, the BCFs are extracted. This process can be independently done by both content owner and data hider. It should be noted that the extraction of the BCFs themselves needs the knowledge of the BCFs. This dilemma is solved by a hierarchically procedure that is described later. Finally, content owner can perfectly reconstruct original image without needing data hider key, $K_d$. As an alternative procedure, we may encrypt the BCFs by $K_s$ before embedding. The key $K_s$ is the third secret key, which can be shared between content owner and data hider.

### A. Analyzing Hiding Capacity

Reference [18] introduces a high capacity RDHEI, in which the prediction error ($e$) is calculated according to

$$p - \rho = e \quad (1)$$

that is difference between a pixel intensity ($p$) and its prediction amount ($\rho$). In [18], it is proved that if the condition

$$|e| < 64 \quad (2)$$

is satisfied, a data bit can be embedded in form of replacing the MSB bit of $p$ by a data bit and at the recovery phase, the MSB of the pixel $p$ can be retrieved without error using $\rho$. The algorithm of [18] embeds at most one data bit in each pixel.

Considering condition (2), we are inspired that the capacity of each pixel may be increased if the condition is replaced as

$$|e| < 2^n \quad 0 \leq n \leq 7 \quad (3)$$

In this way, the embedding capacity of pixel $p$ denoted as $n'$ bits, is given by

$$\begin{cases} n' = 8 - n - 1 & n \neq 0 \\ n' = 8 & n = 0 \end{cases} \quad (4)$$

According to (4), $n' = 7$ never is realized because when $n \neq 0$, a bit should be devoted for $e$ sign and thus, 1 is subtracted to obtain $n'$ in (4). For $n = 0$, (3) gives $e = 0$ and therefore, 8 bits of data can be embedded in $p$. In other words, when the pixel intensity is exactly equal to the prediction one, it can be replaced entirely by the data bits. Regarding $n = 7$, prediction error could be any value in $-128 < e < 128$ range that means there is no capacity to embed data bits. When $n = 6$, just 1 bit capacity ($n' = 1$) is provided as reported in [18]. We describe the data embedding process as

follows in more details:

Let's represent the intensity of a pixel as $p = p_7p_6p_5p_4p_3p_2p_1p_0$ with 8 bits from LSB ($p_0$) to MSB ($p_7$), and express $D = \{d_0, d_1, \ldots, d_{n'-1}\}$ as a set of $n'$ bits of data that should be embedded. The marked pixel, $[\![p]\!]$, that carries $n'$ bits is calculated as

$$[\![p]\!] = \sum_{k=1}^{n'}(2^{8-k} \times d_{k-1}) + \sum_{k=n'+1}^{8}(p_{8-k} \times 2^{8-k}) \quad (5)$$

In this equation, any $\sum_{k=n}^{m}$ with $n > m$ is to be considered as the empty sum, i.e. 0. Thus, a brief description of embedding data in $p$ is that using prediction error, $n$ is calculated ((3)), then the value $n'$ is achieved using (4) and finally, $n'$ bits of the data are embedded in $p$ with respect to (5). As an example, let's assume $|e| < 32$. At first, using (3), $n = 5$ is calculated. Then, $n' = 2$ is attained regarding to $n$ and (4). Finally, using (5), $d_0$ and $d_1$ are embedded in $p$, and we obtain the marked pixel as $[\![p]\!] = d_0d_1\,p_5p_4p_3p_2p_1p_0$.

Knowing $n'$ at the decoding side, it is obvious the most significant $n'$ bits in $[\![p]\!]$ are the embedded bits. It remains to describe how to obtain the original intensity of $p$. First of all, one can obtain the value of $n$ by knowing $n'$ and vice versa. For the case $n = 0$, $p = \rho$ and for $n = 7$, $p$ is not changed in embedding procedure. For $0 < n < 7$, we prove there is only one unique value for $p$ that gives the actual error $e$ and satisfies (3). Let us rewrite Eq. (3) in another form as

$$-2^n < e = (p - \rho) < 2^n \quad (6)$$

When the most significant $n'$ bits of $p$ is embedded with data bits, some of these $n'$ bits and hence, the actual value of $p$ are changed. Any change in the $n'$ bits results in at least $\pm 2^{8-n'}$ change in actual value of $p$, which increases or decreases the error at least $2^{8-n'} = 2^{n+1}$. The error between the marked pixel intensity and prediction one, i.e. $e_n = ([\![p]\!] - \rho)$, will satisfy $e_n > e+2^{n+1}$ or $e_n < e-2^{n+1}$. According to (6), $e_n$ would, therefore, satisfy the conditions $e_n > 2^n$ or $e_n < -2^n$ that is in contradiction with the restriction (6). Consequently, the actual value of $p$ (i.e. the value of $p$ before embedding) can just satisfy the restriction (6).

For the sake of brevity, the image encryption is not described here. The recovery process of the encrypted pixels and extraction data are explained in more detail in Subsection E.

*B. Vacating Room Before Encryption*

In this section, we describe our technique to vacate room before encryption in order to embed data bits after encryption.

In [3], a prediction technique that is local difference of the original image pixels is used to propose a RDH scheme for medical images. This scheme calculates the difference of the neighboring pixels from a base one in separable blocks of the image to construct prediction errors. In our proposed scheme, we introduce a RDHEI that uses the idea of local difference to vacate room before encryption.

Let's assume an original image consists of $M \times L$ pixels and we divide this image to $N_b$ blocks with size of $m \times l$. The permuted version of the blocks is denoted by $B = \{b_1, b_2, \ldots, b_k, \ldots, b_{N_b}\}$. This image blocking may be formed with the sizes of $3 \times 3$, $4 \times 4$ or $5 \times 5$. In these blocks, there is one leader pixel and the rest of the pixels are as followers. So, whole pixels in the image can be categorized in two different sets $P_l$ and $P_f$ that, respectively, present the leader and the follower pixels in all blocks. These sets are denoted as

$$P_l = \{p_{l_1}, p_{l_2}, \ldots, p_{l_k}, \ldots, p_{l_{N_b}}\} \quad (7)$$

$$P_f = \{P_{f_1}, P_{f_2}, \ldots, P_{f_k}, \ldots, P_{f_{N_b}}\} \quad (8)$$

where $p_{l_k}$ denotes one leader pixel and $P_{f_k}$ indicates the set of the follower pixels for the $k$'th block. As an example, Fig. 2 shows the $k$'th $3 \times 3$ block of an image with leader and follower pixels. In this figure, $P_{f_k}(i)$, $1 \leq i \leq 8$ denote 8 follower pixels of $P_{f_k}$ set.

| $P_{f_k}(1)$ | $P_{f_k}(2)$ | $P_{f_k}(3)$ |
|---|---|---|
| $P_{f_k}(4)$ | $p_{l_k}$ | $P_{f_k}(5)$ |
| $P_{f_k}(6)$ | $P_{f_k}(7)$ | $P_{f_k}(8)$ |

Fig. 2. Leader and follower pixels in a $3 \times 3$ block.

| $E_k(1)$ | $E_k(2)$ | $E_k(3)$ |
|---|---|---|
| $E_k(4)$ | $p_{l_k}$ | $E_k(5)$ |
| $E_k(6)$ | $E_k(7)$ | $E_k(8)$ |

Fig. 3. Prediction errors for a $3 \times 3$ block.

There is a correlation between $p_{l_k}$ and $P_{f_k}$ pixels and the local difference between $p_{l_k}$ and $P_{f_k}(i)$ pixel can be viewed as a measure for the correlation. This difference that is called prediction error is given by

$$E_k(i) = P_{f_k}(i) - p_{l_k} \quad 1 \leq i \leq 8 \quad (9)$$

Considering 3×3 block of Fig. 2, prediction error matrix is generated by computing $E_k(i)$ for $1 \leq i \leq 8$ (Fig. 3).

In order to employ the embedding algorithm, described in Subsection A, for a block $k$, we obtain $e_{m_k} = \max(|E_k|)$ and put it in (3) instead of $|e|$ to compute $n$ that is now denoted as $n_k$. In turn, $n'_k$ may be attained from (4) by replacing $n$ with $n_k$. Because the absolute value of any prediction error in a block are not more than $e_{m_k}$, $n'_k$ is the minimum embedding capacity that can be realized for each follower pixel in block $k$. The amount $n'_k$ is considered as the BCF of the block $k$. The number of follower pixels in a block with size of $m \times l$ is, therefore, $(m \times l - 1)$ so that the whole embedding capacity for the block $k$ is given by

$$C_{b_k} = (n'_k) \times (m \times l - 1) \quad (10)$$

Considering the number of blocks in an image is $N_b$, the total embedding capacity will be

$$C = \sum_{k=1}^{k=N_b} C_{b_k} \quad (11)$$

Let's group BCFs of all blocks in set $\{n'_1, n'_2, ..., n'_k, ..., n'_{N_b}\}$. This set is required in decoding phase to extract the embedded data and recover the original image. Given that three bits is required to store each $n'_k$, $1 \leq k \leq N_b$, the number of bits required for BCFs is $3 \times N_b$. After encryption of the permuted image, the BCFs and secret data are embedded using BCFs themselves.

### C. Encrypting the Original Image

The intensity of the $(i,j)$'th pixel is encrypted using bitwise "exclusive or" of the original intensity $O(i,j)$ with a stream cipher as follows.

$$\hat{O}(i,j) = S(i,j) \oplus O(i,j) \quad (12)$$

Set $S$ includes random bytes that are uniformly distributed and may be generated using AES in the CTR mode with an input secret key. It should be noted that classifying the pixels of the original image with regard to (7) and (8), the encrypted image is composed of two different encrypted pixels set as follow

$$\hat{P}_l = \{\hat{p}_{l_1}, \hat{p}_{l_2}, ..., \hat{p}_{l_k}, ..., \hat{p}_{l_{N_b}}\} \quad (13)$$
$$\hat{P}_f = \{\hat{P}_{f_1}, \hat{P}_{f_2}, ..., \hat{P}_{f_k}, ..., \hat{P}_{f_{N_b}}\} \quad (14)$$

Clearly, the original image may be rebuilt using bitwise "exclusive or" operation as

$$O(i,j) = S(i,j) \oplus \hat{O}(i,j) \quad (15)$$

### D. Embedding BCFs and Secret data

In this section, we describe embedding procedure in two steps: 1- embedding the BCFs and 2- embedding the secret data in the encrypted blocks, $\hat{B} = \{\hat{b}_1, \hat{b}_2, ..., \hat{b}_k, ..., \hat{b}_{N_b}\}$. In step 1, we need to initialize hiding process with a block that has significant capacity. Thus, the hiding process is started from a block, $\hat{b}_s$, that has the highest BCF. The address of $\hat{b}_s$, $1 \leq s < N_b$, and $n'_s$ are concatenated and denoted $\mathcal{A}$. For the case of gray scale images, we embed $\mathcal{A}$ in the first three pixels (bytes) of block $\hat{b}_1$. According to any requirement, the number of these pixels can be increased or decreased. In order to retrieve these three pixels at the decoding side, the intensities

**Algorithm 1:** Recovering $P_{f_k}(i)$ from $P'_{f_k}(i)$ and $e'(i)$ for $(1 \leq i < m \times l)$.

---
**for** $i = 1$ **to** $(m \times l - 1)$ **do**
  $e'(i) = P'_{f_k}(i) - p_{l_k}$
  **if** $(n'_k == 8)$ **or** $(n'_k == 0)$ **then**
    $P_{f_k}(i) = P'_{f_k}(i)$
  **else if** $|e'(i)| < 2^{(8-n'_k-1)}$ **then**
    $P_{f_k}(i) = P'_{f_k}(i)$
  **else if** $e'(i) < 0$
    $P_{f_k}(i) = P'_{f_k}(i) + 2^{8-n'_k}$
  **else if** $e'(i) > 0$
    $P_{f_k}(i) = P'_{f_k}(i) - 2^{8-n'_k}$
  **end if**
**end for**

---

of them are included in the data and embedded in step 2.

Starting with $\hat{b}_s$, the BCFs belonging to the next blocks, are embedded in $\hat{b}_s$. After finishing the capacity of $\hat{b}_s$, the next block is used to embed remaining BCFs and the embedding is continued with the other next blocks until all BCFs are finished. $\mathcal{A}$ and BCFs can be encrypted using $K_s$, before embedding.

In step 2, the residual capacity is used to embed the secret data, which is encrypted using $K_d$ before embedding. The embedding algorithm for each block is the one described in Subsection A so that (5) is used to embed data in every $\hat{P}_{f_k}(i)$, $1 \leq i \leq (m \times l - 1)$, where $n'$ and $p$ are replaced by $n'_k$ and $\hat{P}_{f_k}(i)$ respectively. $[\![\hat{B}]\!] = \{[\![\hat{b}_1]\!], [\![\hat{b}_2]\!], ..., [\![\hat{b}_k]\!], ..., [\![\hat{b}_{N_b}]\!]\}$ is the set that represents the marked encrypted blocks.

### E. Extracting Data and Reconstructing the Original Image

In this section, we explain the data bits extraction from marked encrypted blocks and recovering the original image. Having had $n'_k$, i.e. BCFs, the embedded bits in any $[\![\hat{P}_{f_k}(i)]\!]$, $1 \leq i < m \times l$, of $[\![\hat{b}_k]\!]$ are extracted according to

$$D_i^k = \frac{\sum_{i'=1}^{n'_k}(2^{8-i'} \times [\![\hat{P}_{f_k}(i)]\!]_{8-i'})}{2^{8-n'_k}}, \quad 1 \leq i < m \times l, 1 \leq k \leq N_b \quad (16)$$

where $[\![\hat{P}_{f_k}(i)]\!]_q$ is the $q$'th bit of the pixel $[\![\hat{P}_{f_k}(i)]\!]$, for $0 \leq q < 8$.

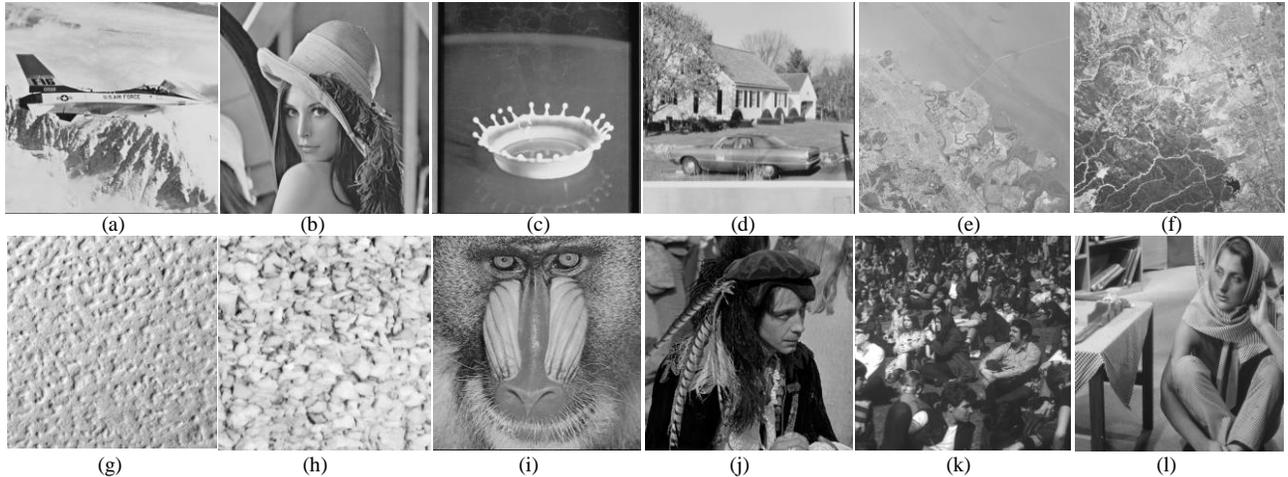

Fig. 4. The test images. (a) F16, (b) Lena, (c) Splash, (d) House, (e) Foster City, (f) Woodland Hills, (g) Rough wall, (h) Gravel, (i) Baboon, (j) Man (k), Crowd and (l) Barbara.

Table I EMBEDDING CAPACITY PROVIDED BY THE PROPOSED ALGORITHM FOR THE TEST IMAGES IN VARIOUS BLOCK SIZES.

| Image blocking | | F16 | Lena | Splash | House | Foster City | Woodland Hills | Rough wall | Gravel | Baboon | Man | Crowd | Barbara |
|---|---|---|---|---|---|---|---|---|---|---|---|---|---|
| $3 \times 3$ | Total capacity (bits) | 809296 | 743448 | 864872 | 711496 | 681352 | 426504 | 448376 | 604544 | 359896 | 598128 | 751168 | 579872 |
| | Secret data (bits) | 721573 | 655725 | 777149 | 623773 | 593629 | 338781 | 360653 | 516821 | 272173 | 510405 | 663445 | 492149 |
| | Secret data (bpp) | 2.75 | 2.5 | 2.96 | 2.38 | 2.26 | 1.29 | 1.38 | 1.97 | 1.04 | 1.95 | 2.53 | 1.88 |
| $4 \times 4$ | Total capacity (bits) | 744420 | 686295 | 810690 | 642405 | 645405 | 349230 | 367005 | 531120 | 298740 | 518820 | 657180 | 521955 |
| | Secret data (bits) | 695268 | 637143 | 761538 | 593253 | 596253 | 300078 | 317853 | 481968 | 249588 | 469668 | 608028 | 472803 |
| | Secret data (bpp) | 2.65 | 2.43 | 2.91 | 2.26 | 2.27 | 1.14 | 1.21 | 1.84 | 0.95 | 1.79 | 2.32 | 1.8 |
| $5 \times 5$ | Total capacity (bits) | 701928 | 642528 | 758640 | 598104 | 624168 | 301608 | 302448 | 470160 | 262176 | 472032 | 586152 | 480672 |
| | Secret data (bits) | 670101 | 610701 | 726813 | 566277 | 592341 | 269781 | 270621 | 438333 | 230349 | 440205 | 554325 | 448845 |
| | Secret data (bpp) | 2.56 | 2.33 | 2.77 | 2.16 | 2.26 | 1.03 | 1.03 | 1.67 | 0.88 | 1.68 | 2.11 | 1.71 |

The extracting procedure is begun by retrieving $\mathcal{A}$ from the first three pixels of $[\![\hat{b}_1]\!]$. $\mathcal{A}$ includes $n'_s$ and address of $[\![\hat{b}_s]\!]$. The BCFs are extracted from $[\![\hat{b}_s]\!]$ using (16) for $k = s$ and it is continued with the blocks after $[\![\hat{b}_s]\!]$ until all BCFs are retrieved. Then, the secret data is extracted using (16) and the retrieved BCFs. The extraction of BCFs may be realized by either content owner or data hider. Lossless reconstruction of the original image may be done by content owner just using his own key ($K_c$) and the BCFs. Therefore, using $K_c$, $[\![\hat{b}_k]\!]$ is decrypted to $[\![b_k]\!]$ that includes $[\![P_{f_k}]\!]$ set and $p_{l_k}$. In order to recover the $i$'th member of $[\![P_{f_k}]\!]$ set (i.e. $[\![P_{f_k}(i)]\!]$), at first, $P'_{f_k}(i)$ is achieved from (17)

$$P'_{f_k}(i) = \sum_{i'=1}^{8-n'_k}(2^{i'-1} \times [\![P_{f_k}(i)]\!]_{i'-1}) + \sum_{i'=9-n'_k}^{8}(2^{i'-1} \times (p_{l_k})_{i'-1}) \quad 1 \le i < m \times l, 1 \le k \le N_b \quad (17)$$

and then, $P_{f_k}(i)$ is restored according to Algorithm 1. In (17), $n'_k$ bits of $p_{l_k}$ that are more significant, are replaced with the corresponding bits of $[\![P_{f_k}(i)]\!]$ to rebuild $P'_{f_k}(i)$.

The recovery process for all blocks of the marked encrypted image is performed according to the aforementioned technique. The reconstructed blocks are permuted using key $k_c$ to reform the original image.

III. EXPERIMENTAL RESULT

We have conducted several experiments to demonstrate the performance of the developed method for embedding data with high capacity in encrypted images. Nine gray scale images F16, Lena, Splash, House, Foster City, Woodland Hills, Rough wall, Gravel and Baboon from the SIPI database and three gray scale images Man, Crowd and Barbara from the miscellaneous database, all 512×512 in size, are used as test images (Fig. 4). Also, BOWS2 original database, including 10000 greyscale images, are exploited to confirm high capacity of the proposed algorithm.

Total embedding capacity and the number of secret data bits, provided by the proposed scheme, are tabulated in Table

Table II PERFORMANCE ANALYSIS OF THE PROPOSED ALGORITHM EMPLOYING THE BOWS2 ORIGINAL DATABASE INCLUDING 10,000 IMAGES.

| Image blocking | Block size = 3 × 3 | | | Block size = 4 × 4 | | | Block size = 5 × 5 | | |
|---|---|---|---|---|---|---|---|---|---|
| | Average | Worst image | Best image | Average | Worst image | Best image | Average | Worst image | Best image |
| Total capacity (bits) | 848685 | 172216 | 1690408 | 790695 | 94560 | 1747095 | 750826 | 55416 | 1788120 |
| Secret data (bits) | 760962 | 84493 | 1602685 | 741543 | 45408 | 1697943 | 718999 | 23589 | 1756293 |
| Secret data (bpp) | 2.9 | 0.32 | 6.11 | 2.83 | 0.17 | 6.48 | 2.74 | 0.01 | 6.70 |
| Image (.pgm) | - | 6184 | 1478 | - | 9448 | 1478 | - | 9448 | 1478 |

Table III COMPARISON OF EMBEDDING CAPACITY OF THE PROPOSED ALGORITHM WITH TWO HIGH CAPACITY RDHEI SCHEMES [18, 19] FOR THE TEST IMAGES.

| images | | F16 | Lena | Splash | House | Foster City | Woodland Hills | Rough wall | Gravel | Baboon | Man | Crowd | Barbara | |
|---|---|---|---|---|---|---|---|---|---|---|---|---|---|---|
| [18] | | 0.98 | 0.98 | 0.99 | 0.94 | 0.99 | 0.97 | 0.99 | 0.99 | 0.75 | 0.92 | 0.98 | 0.76 | bpp |
| [19] | | 2.21 | 2.02 | 2.66 | 1.55 | 1.6 | 0.99 | 1.26 | 1.65 | 0.75 | 1.46 | 1.75 | 1.29 | bpp |
| Proposed scheme | 3 × 3 | 2.75 | 2.5 | 2.96 | 2.38 | 2.26 | 1.29 | 1.38 | 1.97 | 1.04 | 1.95 | 2.53 | 1.88 | bpp |
| | 4 × 4 | 2.65 | 2.43 | 2.91 | 2.26 | 2.27 | 1.14 | 1.21 | 1.84 | 0.95 | 1.79 | 2.32 | 1.8 | bpp |

I, for the test images. They are described for three different sizes of blocks: 3 × 3, 4 × 4 and 5 × 5, that respectively result in 87723, 49152 and 31827 blocks (side blocks are counted for 3 × 3 and 5 × 5 ones). The total embedding capacity is the summation of the BCFs and the secret data bits. Also, the number of secret data bits over the number of pixels in test images, i.e. 512 × 512 pixels, is computed and denoted as bit per pixel (bpp).

By increasing the size of blocks, the number of BCF bits is reduced; however there is more reduction in overall embedding capacity for all test images except for Foster City. Generally, the use of 3 × 3 blocks improves embedding capacity.

Smoother images, such as F16 and Splash, provide more embedding capacity than rougher ones, such as Woodland Hills, Rough Wall and Baboon, regardless of block size. There is better prediction of follower pixels using leader one in smoother images that leads to sharper errors and consequently better embedding capacity.

In Table II, we illustrate the efficiency of the proposed algorithm in all sizes of the block for completely 10000 images of Bows2 original database.

For image blocking with 4 × 4 and 5 × 5 sizes, the best and the worst embedding capacities are provided by image number 1478 and 9448 from Bows2 original database, respectively (Table II). For 3 × 3, the best and the worst ones are numbers 1478 and 6184, respectively. These three Bows2 images are shown in Fig. 5. As tabulated in Table II, image blocking with 5 × 5 size compared to others, improves embedding capacity for the best image. However, for the worst one, employing block with 3 × 3 size increases embedding capacity than others. In average, 3 × 3 blocking provides an improvement more than 0.16 bpp and 0.07 bpp, respectively. In terms of computational complexity, the use of small sizes for blocks increases the number of blocks, BCFs bits, and hence, the computational complexity.

In Table III, the proposed scheme is compared with two high capacity RDHEI schemes [18] and [19] for two different sizes of blocks. The simulation of the cases in [19] is performed for the best α and β, and image blocking with 3 × 3 size. The α and β are two parameters in range $1 \leq \alpha, B \leq 7$ that affect the embedding capacity. Also, the effect of pixels modulation, which reduces slightly embedding capacity, is included in simulation of the method in [19]. This effect can be different in each iteration due to the random nature of the pixel modulation process.

The embedding capacity for one of the two RDHEI schemes proposed in [18] that guarantees perfect reconstruction of the marked encrypted image is also demonstrated in Table III. In this scheme, except for Barbara and Baboon, the other test images indicate embedding capacity near to 1 bpp. However, our proposed scheme gives for all test images more than 1 bpp embedding capacity for 3 × 3 blocking. The proposed method and [19] provide better embedding capacity for smoother images such as F16 and Splash, and less embedding capacity for rougher ones such as Baboon, Rough wall and Woodland Hills.

For all test images and for 3 × 3 size of the image blocking, the proposed algorithm provides better embedding capacity than the scheme in [19]. This improvement for the House and Crowd images is even more than 0.77 bpp. Additionally in our scheme, any standard encryption method or algorithm can be employed to encrypt image.

Fig. 6 demonstrates the PSNR results of recovered images under given hiding capacity in bpp. As clarified in this figure, our scheme achieves more hiding capacity than others in test images, Lena, F16, Man and Crowd. In our method, perfect reconstruction of the original image, i.e. PSNR = ∞, without knowledge of the data hider key is realized while they need both data hider and content owner keys in [16, 20, 21] to absolutely restore original image.

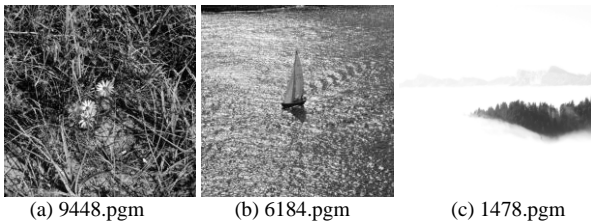

(a) 9448.pgm     (b) 6184.pgm     (c) 1478.pgm

Fig. 5. Three different images from BOWS2 original database. (a) and (b) are the rough images and (c) is a smooth one.

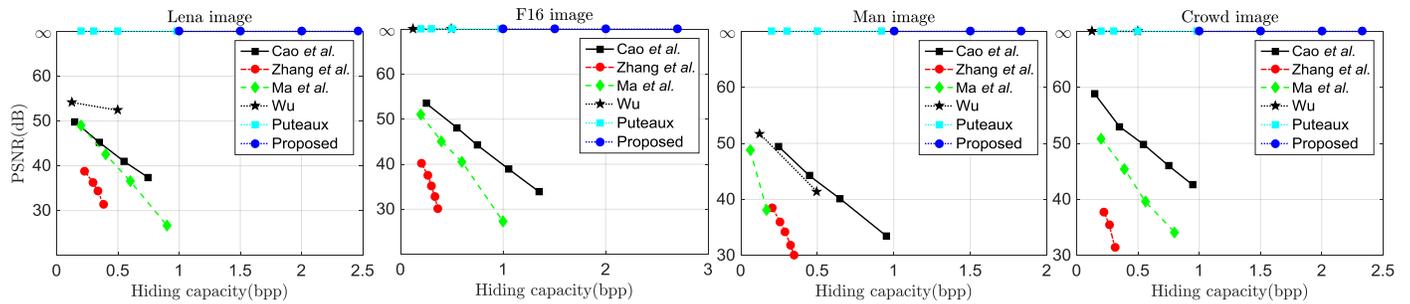

Fig. 6. PSNR comparison between our proposed scheme and other high capacity ones Zhang *et al.* [16], Cao *et al.* [21], Ma *et al.* [20], Wu [22] and Puteaux [18] for four test images.

In [22], even by having both keys, the lossless reconstruction of the original image is possible just for smoother images, like F16 and Crowd. They vacate room after encryption to embed secret data and use based idea of cross-dot predictor [4] to recover original image.

IV. CONCLUSION

In this paper, we propose a high capacity RDHEI that uses the correlation of the pixels in blocks of an image to compute a local difference. The intensity of follower pixels are subtracted from leader one to obtain the local difference or the prediction errors. Analyzing these errors, we bring out the BCFs for the image blocks to vacate rooms before encryption. These features that indicate hiding capacity of the blocks have important role to hierarchically embed and extract data, including themselves and the secret data.

The comparison of the proposed scheme with other state of the art ones confirms that not only image embedding capacity is improved in our scheme, but also content owner can perfectly reconstruct original image without knowledge of data hider key. Also, any standard encryption algorithm can be employed in our developed method.